\documentstyle[aps,twocolumn]{revtex}

\begin{document}
\title{Quantum Cryptography Using Einstein-Podolsky-Rosen Correlations of
Continuous Variables}
\author{Jietai Jing, Qing Pan, Changde Xie and Kunchi Peng}
\address{The State Key Laboratory of Quantum Optics and Quantum Optics\\
Devices,Institute of Opto-Electronics, Shanxi University, Taiyuan, 030006,\\
P.R.China}
\maketitle

\begin{abstract}
Quantum cryptography with the predetermined key was experimentally realized
using Einstein-Podolsky-Rosen(EPR) correlations of continuously bright
optical beams. Only one of two EPR correlated beams is transmitted with the
signals modulated on quadrature phases and amplitudes, and the other one is
retained by the authorized receiver. The modulated small signals are totally
submerged in the large quantum noise of the signal beam, therefore nobody
except the authorized receiver can decode the signals. Usability of
imperfect quantum correlation, high transmission and detection efficiencies,
and security provided by quantum mechanics are the favorable features of the
presented scheme.

PACS numbers: 03.67.Dd,03.67.Hk, 42.50.Dv,89.70.+c
\end{abstract}

Quantum cryptography (QC) provides a way for two distant parties(Alice and
Bob) to share keys for encryption that can be absolutely secret in principle
comparing to the classical ones in which their security is based on the
conjectured computational difficulty of certain functions. The fundamental
properties of quantum mechanics, Heisenberg uncertainty principle, provide
the security of QC. The first complete protocol for quantum key distribution
(QKD) using four polarization states of a photon pulse was proposed in 1984
by Charles Bennett and Gilles Brassard (named BB84)\cite{one}. The majority
of proposals and all implemented experiments for QKD so far have focused on
the use of discrete variables to transmit information\cite
{one,two,three,four,five,six}. Although discrete systems have the advantage
to be loss insensitive, also have a number of disadvantages in working with
single photons, particularly the strong restrictions on data transmission
rates and the poor efficiency of detection. In recent years the
cryptographic schemes employing continuous coherent and nonclassical light
fields have been suggested\cite
{seven,eight,nine,ten,eleven,twelve,thirteen,fourteen,fifteen}. A lot of
interest has arisen in continuous variable quantum cryptography (CVQC) with
EPR beams due to that the experimental demonstrations of quantum
teleportation\cite{sixteen} and quantum dense coding\cite{seventeen,eighteen}
for continuous variables (CV). J.Kimble and his colleagues completed the
quantum communication of dual channels with correlated nonclassical states
of light\cite{thirteen}. However in this case, an eavesdropper (Eve) could
simultaneously access the signal and idler beams without the knowledge of
the legitimate receiver. So far, to the best of our knowledge, there is no
experimental demonstration on CVQC.

In this letter we present a scheme using the nonlocal correlation of bright
EPR beams to implement QC, in which the source generating EPR beams is
placed inside the receiver, only one of the EPR correlated beams (signal
beam) is sent to Alice and other one (idler beam) is retained by Bob. The
security of the presented system is directly provided by EPR correlations
between amplitude and phase quadratures of continuous bright optical beams
and no-cloning of quantum fluctuations. Any random choice on measurements of
quadratures are not needed and the sender encodes the key string with
predetermined binary bits\cite{eleven}. All transmitted bits are used for
constituting the key string without bit rejection, thus the transmission
efficiency of 100\% may be achieved in principle.

It has been theoretically and experimentally demonstrated\cite
{seventeen,eighteen,nineteen,twenty} that bright EPR beams with
anticorrelated amplitude quadratures and correlated phase quadratures may be
produced by a continuous nondegenerate optical parametric amplifier(NOPA)
operating at deamplification. The calculated normalized fluctuation
variances of the amplitude quadratures for the output modes from NOPA are%
\cite{twenty-one} 
\begin{equation}
\left\langle \delta ^2(X_1)\right\rangle =\left\langle \delta
^2(Y_1)\right\rangle =\left\langle \delta ^2(X_2)\right\rangle =\left\langle
\delta ^2(Y_2)\right\rangle =\frac{e^{2r}+e^{-2r}}2
\end{equation}
and the quantum correlation variances between the quadratures are 
\begin{equation}
\left\langle \delta ^2(X_1+X_2)\right\rangle =\left\langle \delta
^2(Y_1-Y_2)\right\rangle =2e^{-2r}
\end{equation}
where $r(0\leq r\leq +\infty )$ is the correlation parameter between modes 1
and 2 which depends on the strength and the time of parametric interaction, $%
r=0$ without correlation, $r>0$ with partial correlation, $r\rightarrow
\infty $ with perfect correlation.

Here we have assumed that the two modes are totally balanced during the
process of measurements. The modulated signals at given radio frequencies
(rf) can be considered as a noise $\delta X_S(\Omega )$ and $\delta
Y_S(\Omega )$. When the powers of $\delta X_S(\Omega )$ and $\delta
Y_S(\Omega )$ are smaller than the original quantum noise power of the
signal beam (Eq.(1)) and larger than the correlation noise power (Eq.(2)),
the signals are submerged in the noise background of the signal beam $%
(X_1,Y_1)$ and can be decoded with the correspondent idler beam $(X_2,Y_2)$
of the EPR beams. Thus, the strength $\left\langle \delta ^2X_S\right\rangle 
$ of the modulated signals should satisfy the following inequalities: 
\begin{eqnarray}
2e^{-2r(\Omega )} &<&\left\langle \delta ^2X_S(\Omega )\right\rangle <\frac{%
e^{2r(\Omega )}+e^{-2r(\Omega )}}2 \\
2e^{-2r(\Omega )} &<&\left\langle \delta ^2Y_S(\Omega )\right\rangle <\frac{%
e^{2r(\Omega )}+e^{-2r(\Omega )}}2  \nonumber
\end{eqnarray}
From Eqs.(1) and (2), it is obvious, when $r>0.27$ the variances of signal
beam(Eq.(1)) are higher than that of the correlation fluctuations(Eq.(2)).
The larger the correlation parameter is, the higher the noise power of the
signal beam is and the lower the correlation variance is. For a perfect
transmission line of noiseless the correlation parameter of the NOPA should
be larger than $0.27$ at least for accomplishing the QC. At Bob station, the
signals $X_S(\Omega )$ and $Y_S(\Omega )$ modulated on the amplitude and
phase quadratures of signal beam $\left( X_1,Y_1\right) $ are decoded with
the retained idler beam by means of the direct detection of photocurrents 
\cite{seventeen,eighteen}.

Fig.1 is the schematic of the QC system. The EPR source, NOPA, is placed
inside the Bob receiving station. The one of the bright EPR correlated
beams, the signal beam $\left( X_1,Y_1\right) $, is sent to the Alice
sending station where Alice encodes the transmitted information on the
amplitude and phase quadratures by the choice of the modulation types with
the binary bit values, for example the modulated amplitude signals stands
for $``1"$and the phase signals for $``0"$. Then the encoded signal beam is
transmitted back to Bob where the information is decoded by the retained
other one of the EPR correlated beams.

The configuration of NOPA and scheme producing the bright beams with
anticorrelation of amplitude quadratures and correlation of phase
quadratures were same as that used in the experiment of the quantum dense
coding which was described in detail in our previous publication\cite
{seventeen}. Locking the relative phase between the pump laser and injected
seed wave of NOPA to $(2n+1)\pi $, where $n$ is integer, to enforce the NOPA
operating at deamplification, the entangled EPR beams were generated. The
output field from NOPA is splitted into the signal and idler beams by a
polarized-beam-splitter (PBS), then the signal beam is sent to Alice and the
idler beam is retained by Bob. The phase shifter (PS), the 50-50 beam
splitter (BS), a pair of photodiodes D1 and D2, and the power combiners
(+/-) in Bob station constitute the direct detection system of Bell-state
for measuring the variances of the amplitude sum and phase difference
between the two quadratures simultaneously\cite{seventeen,eighteen}. The PS
imposes a phase difference of $\pi /2$ between signal and idler beams before
they are mixed on BS\cite{eighteen}.

A secretly encoded and transmitted binary bit string is summarized in Table
I. At the time points $t_1,t_2...t_6$, Alice encodes the signal beam and
sends it back to Bob, then Bob simultaneously measures the amplitude and
phase modulation signals. The measured amplitude signals at $t_1,t_4,t_5$
stand for $``1"$, the phase singles at $t_2,t_3,t_6$ for $``0"$. The upper
traces are the spectral densities of noise power of the
SNL(Shot-Noise-Limit) for the EPR beams which is used to evaluate the
quantum correlations. The middle traces are the noise spectra of signal beam
in which the modulated signals are submerged in the noise background totally
and the lower traces are the variances $\left\langle \delta
^2(X_1+X_2)\right\rangle $ and $\left\langle \delta ^2(Y_1-Y_2)\right\rangle 
$ , which are 3.8dBm below the noise level of the SNL of the EPR beams, and
the secret signals modulated at 2MHz emerge from the squeezed noise
backgrounds. A secret key string characterized by the binary bit values
100110 is obtained.

In experiments, we should pay attention on matching the optical distances
between the signal and idler beams to reach optimum correlation and
detection efficiency. Especially for long distance transmission one must add
optical retarder and attenuator in the idler beam retained by Bob to make
its travel distance from EPR source to the beam splitter of the detection
system and the detected intensity approximately equal to that of the signal
beam. Recently, the schemes on storing quantum entanglement in atomic spins
system or in cavity QED have been presented\cite{twenty-two,twenty-three},
that probably can provide a practical technique for retaining an EPR
component in Bob.

The intensity fluctuations of signal and idler fields of the bright EPR
beams generated by the NOPA are quantum-correlated at any instant and nobody
can make or copy a light field which has the identical quantum fluctuations,
which provide the mechanism of security for the CVQC system. In experiment,
the SNR(signal to noise ratio) of the decoded signals depends on the
original correlation of EPR beams and the losses of a given transmission
line, which is independent of Alice's bit value. Any decrease of the
measured SNR and the correlation degree relative to the predicted results
alerts Bob to the additional loss caused by a partial tapping of the signal
channel, perhaps by Eve. If Eve wants to perform the
quantum-nondemonlition-measurment (QND) on one of the amplitude-phase
quadratures of the signal beam, the other quadrature must be disturbed and
the disturbance must be reflected on the measured results of Bob. The
deviation would indicate the possibility that Eve have performed a QND. More
quantitative analyses on protecting against the optical tap attack of Eve
have been presented in the previous publications\cite
{ten,eleven,twelve,thirteen} and can be used in the discussion to the
presented protocol.

A possible eavesdropping scheme using fake EPR beams is shown in Fig.2. Eve
intercepts totally the signal beam $(X_1,Y_1)$ and transmits a fake signal
beam $(X_1^{^{\prime }},Y_1^{^{\prime }})$ produced by a fake EPR source$%
(EPR2)$ in her station to Alice. Alice has no ability to recognize the fake
beam. She modulates the information on it and then sends out as usual. Eve
intercepts the beam with messages again and decodes the information with the
other one of the fake EPR correlated beams $(X_2^{^{\prime }},Y_2^{^{\prime
}})$ retained by her. At the same time she modulates the real signal beam $%
(X_1,Y_1)$ according to the intercepted bit values and sends it back to Bob
who also does not know that the information has been intercepted. To reveal
this type of quantum interception Alice may randomly block the signal beam
with a photoelectric detector ($D_o$) connected to an oscilloscope at some
appointed time points $t_k$ for a short time interval $\Delta t_k$ during
the duration of transmission. The shape of intensity fluctuation of the
signal beam at the blocked moment can be recorded as a function of time by
an oscilloscope ($OS_1$). In Bob station, the photocurrent of the sum of the
amplitude quadratures $\left\langle \delta ^2(X_1+X_2)\right\rangle $ is
splitted to two parts by a power splitter, one is sent to a spectrum
analyzer (SA) for the noise spectrum measurement and other one to an
oscilloscope ($OS_2$) for recording the fluctuation of photocurrents. If
there is no Eve between Alice and Bob, while the signal beam is blocked by $%
D_o$ the oscilloscope $OS_2$ in Bob station records the intensity
fluctuation of the idler beam $\left\langle \delta ^2X_2(t)\right\rangle $
which is anticorrelated with $\left\langle \delta ^2X_1(t)\right\rangle $
recorded by $OS_1$ at same time and the correlation extent only depends on
the quality of the EPR source (EPR1) and the losses of transmission line.
Fig.3 shows the intensity fluctuation shapes of $\left\langle \delta
^2X_1(t)\right\rangle $ (trace1) and $\left\langle \delta
^2X_2(t)\right\rangle $ (trace2) simultaneously recorded by $OS_1$ and $OS_2$
respectively. The partial anticorrelation between the shapes of the
fluctuations is very obvious and the measured root-mean-square (rms) noise
voltages for the sum and difference of traces 1 and 2 are $568.64\mu v$ and $%
655.29\mu v$ respectively. For two uncorrelated beams, the rms voltages for
sum and difference photocurrents should be equal, thus the deviation between
rms voltages of the sum and difference also shows the presence of quantum
correlation. If Eve interrupts the real signal beam, the intensity
fluctuations of $\left\langle \delta ^2X_1^{^{\prime }}(t)\right\rangle $
recorded by Alice at the blocked moment are totally not correlated with that
recorded by $OS_2$ at same time. After a set of communication is finished,
Alice publicly sends the wave shapes of intensity fluctuation and the rms
voltages recorded by $OS_1$ to Bob, where Bob compares with that recorded by
himself with $OS_2$ at same time points. The absence of anticorrelation
between the two sets of fluctuation shapes and the same rms voltages for the
sum and difference indicate the presence of Eve.

At last we have to mention that in the practical communication of long
distance the question of preserving the entanglement of the initial EPR
state is a fascinating one for continuous quantum variables due to the
inevitable losses in transmission lines and the sensitivity of the quantum
correlation to losses. However the usability of experimentally accessible
quantum correlation, relatively high security and efficiency, and the
simplicity of the configuration make this scheme valuable to be considered
for developing CVQC.

Acknowledgements. This work was funded by the National Fundamental Research
Program (No.2001CB309304) and the National Nature Science Foundation
(No.69837010). We are grateful to J. H. Kimble, Min Xiao, Jing Zhang and
Jiangrui Gao for helpful discussions.

Captions of figures:

Fig.1 The schematic of the quantum cryptography using the EPR beams

AM-amplitude modulator, PM-phase modulator, RNG-random number generator,
SA-spectrum analyzer, PS-phase shifter, BS-beamsplitter of 50\%,
PBS-polarization beamsplitter, NOPA-nondegenerate optical parametric
amplifier.

Fig.2 Diagram of eavesdropping Scheme using fake EPR beams

EPR1, EPR2-Sources of EPR beams, AM, PM-Amplitude and phase modulators,
SA-spectral analyzer, $OS_1,OS_2$-Oscilloscopes, Detector1, 2-Detection
systems of Bell-state, $D_o$-photoelectric detector, RNG-random number
generator.

Fig.3 The fluctuation shapes of signal (1) and idler (2) beams recorded by
the oscilloscopes OS1 and OS2 at Alice and Bob.

Table I The generation of the predetermined secret key string

AQ-Amplitude modulation signals, PQ-Phase modulation signals,CD-The
correlation degrees between signal and idler beams measured by Bob, Key-The
secret keys, upper traces-The SNL of EPR beams, middle traces-the noise
spectra of signal beam, lower traces-the noise spectra of $X_1+X_2$ and $%
Y_1-Y_2.$The measured frequency range 1.0-3.0MHz, resolution bandwidth
30KHz, video bandwidth 0.1KHz. The electronic noise is about 8dBm below that
of the SNL of the EPR beams.(not indicated in the figures)


\begin{references}
\bibitem{one}  C. H. Bennett and G. Brassard, Proc.IEEE Int. Conf. On
Computers, Systems and Signal Processing(Bangalore), 175 (1984)

\bibitem{two}  W. Tittel, J. Brendel, H. Zbinden, and N. Gisin, Phys. Rev.
Lett. 84(20), 4737 (2000)

\bibitem{three}  T. Jennewein, C. Simon, G. Weihs, H. Weinfurter, and A.
Zeilinger, Phys. Rev. Lett. 84(20), 4729 (2000)

\bibitem{four}  D. S. Naik, C. G. Peterson, A. G. White, A. J. Berglund, and
P. G. Kwiat, Phys. Rev. Lett. 84(20), 4733 (2000)

\bibitem{five}  W. T. Buttler, R. J. Hughes, S. K. Lamoreaux, G. L. Morgan,
J. E. Nordholt, and C. G. Peterson, Phys. Rev. Lett. 84(24), 5652 (2000)

\bibitem{six}  G. Rigbordy, J. D. Gautier, N. Gisin, O. Guinnard, and H.
Zbinden, J. Mod. Opt. 47, 517 (2000)

\bibitem{seven}  D. Gottesman and J. Preskill, Phys. Rev. A 63, 022309 (2001)

\bibitem{eight}  M. Hillery, Phys. Rev. A 61, 022309 (2000)

\bibitem{nine}  N. J. Cerf, M. Levy and G. Van Assche, Phys. Rev. A
63,052311 (2000)

\bibitem{ten}  T. C. Ralph, Phys. Rev. A 62, 062306 (2000)

\bibitem{eleven}  M. D. Reid, Phys. Rev. A 62, 062308 (2000)

\bibitem{twelve}  Ch. Silberhorn, N. Korolkova and G. Leuchs,
quant-ph/0109009 (2001)

\bibitem{thirteen}  S. F. Pereira, Z. Y. Ou and H. J. Kimble, Phys. Rev. A
62, 042311 (2000); H. J. Kimble, Z. Y. Ou, and S. F. Pereira, U.S. Patent
No. 5,339,182, Issued 8/16/94

\bibitem{fourteen}  K. Bencheikh et al., J. Mod. Opt. 48, 1903 (2001)

\bibitem{fifteen}  F. Grosshans and P. Grangier, Phys. Rev. Lett. 88, 057902
(2002)

\bibitem{sixteen}  A. Furusawa, J. L. Sorensen, S. L. Braunstein, C. A.
Fuchs, H. J. Kimble, E. S. Polzik, Science 282,706 (1998)

\bibitem{seventeen}  X. Li, Q. Pan, J. Jing, J. Zhang, C. Xie, K. Peng,
Phys. Rev. Lett. 88, 047904 (2002)

\bibitem{eighteen}  J. Zhang and K. Peng, Phys. Rev. A 62, 064302 (2000)

\bibitem{nineteen}  M. D. Reid and P. D. Drummond, Phys. Rev. Lett. 60,
2731(1988)

\bibitem{twenty}  Y. Zhang, H. Wang, X. Li, J. Jing, C. Xie, K. Peng, Phys.
Rev. A 62, 023813 (2000)

\bibitem{twenty-one}  Y. Zhang, H. Su, C. Xie and K. Peng, Phys. Lett. A
259, 171 (1999)

\bibitem{twenty-two}  B. Julsgaard, A. Kozhekin and E. S. Polzik, Nature
413, 400 (2001)

\bibitem{twenty-three}  A. S. Parkins and H. J. Kimble, Phys. Rev. A 61,
052104(2000)
\end{references}
\end{document}